\documentstyle[11pt,aaspp]{article}
\input{psfig}
\eqsecnum
\def\beq{\begin{equation}}
\def\eeq{\end{equation}}
\def\ref{\reference}
\def\simge{\mathrel{%
   \rlap{\raise 0.511ex \hbox{$>$}}{\lower 0.511ex \hbox{$\sim$}}}}
\def\simle{\mathrel{
   \rlap{\raise 0.511ex \hbox{$<$}}{\lower 0.511ex \hbox{$\sim$}}}}

\begin{document}

\title{An Explanation for the Bimodal Duration Distribution of Gamma-Ray
Bursts: Millisecond Pulsars from Accretion-Induced Collapse}

\author{Insu Yi$^1$ and Eric G. Blackman$^2$}

\affil{$^1$Institute for Advanced Study, Olden Lane, Princeton, NJ 08540; 
yi@sns.ias.edu}
\affil{$^2$Institute of Astronomy, Madingley Road, Cambridge CB3 OHA, England;
blackman@ast.cam.ac.uk}

\vskip 0.3cm

\begin{abstract}

Cosmological gamma-ray bursts (GRBs) could be driven by dissipation 
of pure electromagnetic energy (Poynting flux) extracted from rapidly 
rotating compact objects with strong magnetic fields. One such possibility 
is a young millisecond pulsar (MSP) formed from 
accretion-induced collapse (AIC) of a white dwarf.  
The combination of an efficient magnetic dynamo, likely 
operating during the first seconds of the initially hot and turbulent MSP 
interior, and the subsequent modest beaming of gamma-ray emitting outflows, 
would easily account for energy constraints. But the 
remarkable feature of such models is that they may naturally explain 
the hitherto unexplained bimodal distribution in GRB time durations.  
The two burst classes could correspond to MSPs
that form spinning above and below a gravitationally unstable limit 
respectively. In the former case, the spin-down time scale is due to 
gravitational radiation emission ($<1s$) while the spin-down time scale 
of the latter is due to electromagnetic dipole emission ($\gg 1s$). These 
two time scales account for the short and long GRB durations, i.e. the 
observed bimodal GRB duration distribution. 
A natural prediction is that the short duration GRBs would be 
accompanied by strong gravitational 
radiation emission which is absent from the longer class.
Both would show millisecond variabilities.

\end{abstract}


\section{Introduction}

If the recently observed GRB afterglows in optical and radio are due to
an expanding fireball, the total gamma-ray burst energy output is 
likely to be $\sim 10^{51}\Delta\Omega/4\pi~erg$ where $\Delta\Omega$ 
is the solid angle for the opening of the outflow from which emission occurs
(e.g. Waxman et al. 1997, Dar 1997, and references therein).
The fireball and its afterglow represent a generic description of how
the GRB engine energy is dissipated, but there are a number
of possibilities for what actually powers the engine.
The well-known neutrino fireball model in the context of the neutron star
mergers does not seem to provide sufficient energy to power the extended
afterglow phase lasting $\sim months$ unless some beaming is present (Dar 1997).
The required gamma-ray transparency condition on the bulk
Lorentz factor of the outflow ($\Gamma>100$) severely limits its
baryon fraction (e.g. Fenimore et al. 1993).
These problems have led to models in which GRBs are powered by rapidly
spinning compact objects with strong magnetic fields 
(e.g. Usov 1992, Blackman et al. 1996, Meszaros \& Rees 1997, Paczynski 1997). 
In such models, power is extracted in the form of the pure 
electromagnetic low-frequency waves (Poynting flux) which are later converted 
to gamma-rays and lower frequency photons (Usov 1992, 
Blackman et al. 1996, Meszaros \& Rees 1997). 
Regardless of model details, the central engines could be  
expected to possess rapid rotation with spin frequencies $\sim 10^4 s^{-1}$ 
and strong magnetic fields $\sim 10^{15}G$ (Usov 1992, Meszaros \& Rees 1997).
Beaming should somehow be plausible for such rapidly rotating systems. 

The GRB engines could  either be rapidly rotating black holes surrounded by 
strongly magnetized tori or rapidly rotating MSPs
with strong magnetic fields (Usov 1992, Meszaros \& Rees 1997).
In the former scenario, the black holes could form as a consequence
of the neutron star-neutron star mergers and the strong fields could be
generated by dynamos in the rapidly rotating tori which are direct merger 
remnants. In the latter case, as
first proposed by Usov (1992), the MSPs could form from the accretion induced
collapse of white dwarfs. Yi and Blackman (1997) have shown
that typical AIC events cannot produce a MSP with a magnetic field as strong 
as $\sim 10^{15} G$ if the field is solely due to the flux-frozen white dwarf 
fossil field. However, strong magnetic fields $\sim 10^{15}G$ could be 
generated by dynamo action during the hot, convective phase of the MSPs which 
directly follows the AIC (Duncan \& Thompson 1992).

Despite such developments, it remains a challenge to account for
the vastly different time scales seen in various types of GRBs (e.g.
Fishman \& Meegan 1995).  Short variability time scales range $\sim 10^{-3}s$ 
to $\sim 0.1s$ with a myriad of complex time profiles, but
most notably, the GRB durations are known
to be distributed bimodally. The short durations range
from $\sim 3\times 10^{-2} s$ to $\sim 2s$ while the long durations
range from $\sim 2s$ up to $\sim 10^3s$. 

The bimodal distribution suggests that something discrete 
distinguishes the two classes.  
An interesting advantage of the AIC/MSP scenario over the black
hole scenario in this respect is that 
two classes of pulsars naturally form (Usov 1992), 
and can therefore explain the bimodality.
When a MSP rotates with spin frequency larger than a certain critical
frequency (e.g. Friedman 1983), non-axisymmetric secular instability drives 
the pulsar into non-axisymmetric configuration with non-zero quadrupole moment.
In this case, the spin-down of the pulsar occurs on a time scale 
\beq
\tau_{gw}=I_{*}\Omega_{*}^2/2L_{gw}\sim 3\times 10^{-3} \epsilon^{-2}
I_{*,45}\Omega_{*,4}^{-4}~~ s
\eeq
where $I_{*}=I_{*,45} 10^{45} g cm^2$ is the MSP moment of inertia
$\epsilon=\delta R_*/R_*$ is the ellipticity of non-axisymmetrically 
deformed pulsar with radius $R_*$ and perturbation $\delta R_*$, 
$\Omega_{*}=\Omega_{*,4}10^4 s^{-1}$ is the MSP spin 
frequency, and we have made use of the gravitational wave energy loss rate
\beq
L_{gw}=32G\epsilon^2I_{*}^2\Omega_{*}^6/5c^5.
\eeq
On the other hand, the spin-down due to the electromagnetic dipole radiation
gives a time scale
\beq
\tau_{em}\sim I_{*}\Omega_{*}^2/2L_{em}\sim 3\times 10^2 I_{*,45}
R_{*,6}^{-6} B_{*,15}^{-2} \Omega_{*,4}^{-2}~~ s
\eeq
where $R_{*}=R_{*,6}10^6cm$ is the MSP radius, $B_{*}=B_{*,15}
10^{15} G$ is the MSP dipole magnetic field and we have assumed the
simple electromagnetic dipole energy loss rate
\beq
L_{em}=2\mu_{*}^2\Omega_{*}^4/3c^3
\eeq
where $\mu_{*}=B_{*}R_{*}^3$ is the electromagnetic dipole moment.
Usov (1992) pointed out these two natural time scales, and here
we explore how they may account for the 
bimodal distribution of GRBs. 

We consider the AIC scenario and carefully examine how
plausible physical conditions may facilitate an explanation for bimodality.
It turns out that it is necessary to consider hot neutron 
star conditions under which high neutrino viscosity and convection are 
important as noted by Duncan and Thompson (1992). 
We then discuss possible 
implications of the AIC-MSP interpretation of the bimodal duration
distribution.

\section{Pulsars near Critical Rotation}

When a white dwarf reaches the critical Chandrasekhar mass $\sim 1.4M_{\odot}$
through mass accretion, the white dwarf can collapse to a neutron star.
The probable mass accretion rate leading to AIC is ${\dot M}\simge 3\times
10^{18} g/s$ (e.g. Livio \& Truran 1992). Even at these high mass accretion
rates, the mass accretion should be sustained for $>10^6 yr$. Such constraints
generally disfavor dwarf binary systems in which the secondary mass
donors are dwarf stars. The white dwarf magnetic fields interact with
accretion flows and directly affect its subsequent spin evolution. 
But given the long accretion time scales, the initial
spins of the white dwarfs have little effect on the final outcome of
the pre-collapse white dwarf (Yi \& Blackman 1997).
For a white dwarf with  moment of inertia
$I_{wd}=10^{51} g cm^2$ and  radius $R_{wd}=10^9 cm$, the magnetic
field $B_{wd}$ and spin frequency $\Omega_{wd}$
are related to the post-AIC pulsar spin frequency and magnetic field
by $\Omega_{wd}=\Omega_{*}(I_{*}/I_{wd})$ and
$B_{wd}=B_{*}(R_{*}/R_{wd})^2$ where $I_{*}=10^{45} g cm^2$
and $R_{*}=10^6 cm$ are the moment of inertia and the radius of the
MSP. In order to create a MSP ($\Omega_{*}\sim 10^4 s^{-1}$) with
$B_*\sim 10^{15} G$ by flux-freezing, 
the pre-collapse white dwarf must have
$\Omega_{wd}\sim 10^{-2} s$ and $B_{wd}\sim 10^9 G$. Such strong
white dwarf fields have not been observed. 
However, even when such a field exists, Yi \& Blackman (1997) 
have shown that such pre-AIC magnetized accretion is not compatible with 
such a white dwarf due to efficient magnetic braking and spin-down
during pre-AIC magnetized accretion phase. They find that
the most likely AIC-produced MSP parameters satisfy
\beq
\Omega_{*}\sim 10^4 B_{*,11}^{-4/5} s^{-1}
\eeq
where $B_{*,11}=B_{*}/10^{11}G$. 

Duncan \& Thopmson (1992) suggested that the hot MSP
which is likely to form from AIC is a favorable site for an efficient
dynamo of $\alpha\omega$ type due to the vigorous convection driven
by a large neutrino flux which is the direct outcome of the white
dwarf to neutron star collapse. During the first seconds, the large
neutrino flux drives convection.
When the convection is significant, the efficiency of the dynamo
could be roughly estimated by the Rossby number 
\beq
N_{R}=P_{*}/\tau_{conv}
\eeq
where $P_{*}=2\pi/\Omega_{*}$ is the MSP spin period
and $\tau_{conv}$ is the convective overturn time scale at the base of
the convection zone. If $N_R$ is of order unity or less in a 
turbulent medium, the amplification of field by helical motion is
not suppressed by turbulent diffusion and an efficient dynamo results.
During the hot neutrino phase, the neutrino viscosity greatly
exceeds the kinematic shear viscosity appropriate to cool
neutron stars, $\nu\sim 1-100 cm^2 s^{-1}$.
Following Duncan and Thompson (1992),
the convective overturn time is roughly given by $\tau_{conv}\sim 
10^{-3} F_{39}^{-1/3} s$ where $F_{39}$ is the convective neutrino
heat flux in units of $10^{39} erg/s/cm^2$.

The uncertainty associated with the required spin rate for an efficient dynamo 
action, $\Omega_{dynamo}\sim 2\pi \tau_{conv}^{-1}\simle 6\times 10^3 s^{-1}$ 
is largely due to the uncertainty in the effectiveness of the dynamo when 
$N_{R}\sim 1$ and the uncertainty in the convective overturn time scale.
The increase of $N_R$ from $\sim 1$ by an order of magnitude seems to quench
the build-up of the strong fields (Simon 1990). Nevertheless, 
Duncan \& Thompson (1992) show that a dynamo generated large scale dipole 
field $\sim 10^{15}G$ can result. Given the possibility of some dynamo action 
at $N_R\simge 1$, the dynamo-generated field may still exist for 
$\Omega_*\simle \Omega_{dynamo}$.

The secular instability driven by the gravitational wave emission
and damped by the shear viscosity perturbs the axi-symmetric star
with a non-axisymmetric perturbation of the form
(e.g. Wagoner 1984)
\beq
\delta R_*=\Sigma_{l,m}\delta R_{lm}Y_{l}^m(\theta,\phi)\exp(i\omega t)
\eeq
where the instability sets in via a mode with $\omega=m\Omega_*$
(Friedman 1983). There are in principle two ways in which a gravitationally
unstable star can radiate away energy.  It can 1) spin down, or 
2) spin up, but lower its moment of inertia accordingly.
The former is relevant when the fast spinning neutron star
is approximated as a classical Maclaurin spheroid. 
This approximation amounts to the condition that
the neutron star equation of state is sufficiently stiff
(e.g. Chandrasekhar 1969, Yi \& Blackman 1997) and that 
the internal vorticity of the star in the rotating ellipsoidal pattern
frame, exceeds twice the pattern speed measured in the inertial frame
(Lai \& Shapiro 1995).  
Evolution 2) corresponds to a Jacobi to Maclaurin transition, and
would occur when the inequality, described above, is reversed.
Track 2) also requires the gravitational perturbation time to 
be slower than the gravitational radiation time, otherwise
the star could not adjust its radius quickly enough to change its moment
of inertia as required.  Though the initial conditions determine
which evolution the star will follow, we will see that the growth
time turns out to be of order the gravitational radiation time.
We therefore suspect that spin-down evolution is generally more likely,
and we focus on the Maclaurin evolution.  

The perturbation growth time scale for the Maclaurin
spheroid is
\beq
\tau_{gr}={(m-1)[(2m+1)!!]^2\over (m+1)(m+2)(1-e^2)^{1/2}}\left(c\over
R_*\omega\right)^{2m+1}\left(\omega+(m-1)\Omega_*\over 2\pi G\rho_*\right)
\eeq
where $\rho_*$ is the stellar density and $e$ is the eccentricity
of the MSP when the instability sets in (Comins 1979, Friedman 1983,
Lindblom 1986). For a neutron
star $t_{dyn}^{-1}=\sqrt{\pi G\rho_*}\approx 10^4 s^{-1}$. 
Since $\omega \sim t_{dyn}^{-1}$ using the values of $e$ for each model $l=m$,
the growth time scale becomes (Friedman 1983)
\beq
\tau_{gr}\approx a t_{dyn}(c/R_*\omega)^{2m+1}
\eeq
where $a\approx 10, 10^3, 10^5, 10^7$ for $m=2,3,4,5$ respectively. Therefore,
$\tau_{gr}\approx 1, 10^3, 10^6, 10^9 s$ for $m=2,3,4,5$. The major uncertainty
comes from the sensitive dependence of $\tau_{gr}$ on $\omega$.
The $l\neq m$ modes become unstable only after $l=m$ bar modes become
unstable so they are irrelevant for our discussions.

The large neutrino viscosity of hot young pulsars leads to 
much shorter damping time scales compared to those of cold neutron stars. 
The viscous damping time scale is (Comins 1979, Friedman 1983, Lindblom 1986)
\beq
\tau_{vis}\approx {1\over (m-1)(2m+1)}{R_*^2\over \nu}
\eeq
In ordinary cold neutron stars, $\nu\sim 1-100 cm^2/s$ and hence the
viscous damping time scale $\tau_{vis}\approx [(10^{10}-10^{12})/(m-1)(2m+1)]s$.
All modes with $m\le 4$ are in principle excited. The critical frequency
is then set by $m=4$ mode as the highest $m$ mode gives the lowest
critical frequency. 
However, right after AIC, when MSP is hot, the kinematic shear viscosity
is determined by the neutrino viscosity which is close to $\sim$ a few $\times
10^9 cm^2 s^{-1}$ for $\rho_*=10^{15} g/cm^3$ and the MSP temperature of
$\sim 10^{10} K$ (Duncan \& Thompson 1992). 
The corresponding viscous damping time scale is
$\tau_{vis}\approx [10^3/(m-1)(2m+1)]s$ or $\tau_{vis}\approx 2\times
10^2, 70, 40, 20 s$ for $m=2,3,4,5$, respectively. Therefore, for this high
viscosity, all modes with $m>2$ are rapidly damped but the $l=m=2$ mode
grows. The critical frequency for the secular instability is
then simply determined by the $l=m=2$ instability. Despite various
uncertainties in estimating the time scales, $\tau_{gr}<\tau_{vis}$ is
likely to be satisfied by the $l=m=2$ mode.
This implies that the non-axisymmetric perturbation occurs at
a frequency closest to the dynamical break-up frequency. 
In cool neutron stars (with low viscosity), 
the critical frequency for the secular instability is determined
by higher $l=m$ modes and as a consequence the 
critical frequency is substantially lower than that associated
with the  $l=m=2$ mode.

For $l=m=2$ mode in the Maclaurin spheroids, the critical frequency
is given by (Chandrasekhar 1969)
\beq
\Omega_{crit}=0.612t_{dyn}^{-1}
\eeq
at which point the eccentricity of the spheroid $e=0.813$.
The effect of the relativistic corrections is at the level of $\sim
10\%$ if $GM/Rc^2\sim 0.3$ (Balbinski \& Schutz 1982). 
For pulsars with $\Omega_{*}>\Omega_{crit}$, the $l=m=2$ model leads to
perturbed ellipsoid and develops a non-zero quadrupole moment. The spin-down
during this stage is roughly described by
\beq
{d\Omega_{*}\over dt}=-{s\Omega_{crit}\tau_{gr}^{-1}}{M_*(\delta R_*)^2\over
I_{*}},
\eeq
where $s$ is a constant of order unity. 
The  $l=m=2$ mode grows quickly, building 
up to $\delta R_*\simle R_*$.  Further growth is limited by nonlinear
effects. 

The loss of energy by 
gravitational radiation from an unstable Maclaurin spheroid
proceeds through spin down.
The gravitational radiation depletes energy and 
angular momentum while $\delta R_*$ remains nearly constant. 
Eventually, the gravitational instability growth 
time scale gets longer as spin-down continues until $\tau_{gr}>\tau_{vis}$
occurs for $l=m=2$. For $s\sim 1$, $I_{*}\sim constant$, $\delta R_*\sim R_*$
and $\Omega_{*}\sim \Omega_{crit}$, the spin-down time scale
$\tau_{down}\sim \tau_{gr}$.

In general, when the gravitational instability growth time scale is of order
the gravitational radiation drain time scale the star does not
have time to change its moment of inertia, so gravitational radiation proceeds
through spin-down.  This is a likely state for a highly viscous
nascent neutron star.  This also 
means that the electromagnetic luminosity is
depleted on the spin-down time scale,
enabling the AIC-MSP model to 
account for the bimodality of GRB durations, as addressed in the next
section.  Note that if the young NS initially were somehow to 
follow the Jacobi track 
2) as discussed above, it would instead spin up, and would not
naturally lead to a bimodal distribution.  However, the gravitational
radiation from these two paths are very distinct, and will be
measurable (Lai \& Shapiro 1995).  Either way, the signature
of GRB from AIC-MSP should be testable.

\section{Bimodal Distribution}

Based on the discussion of the previous section, we classify three
classes of hot MSPs relating to GRB engines.
This essentially involves three rotational frequencies
$\Omega_{*}$, $\Omega_{dynamo}$, $\Omega_{crit}$ (cf. Blackman et al.
1996). If MSPs are formed from AIC of magnetized white dwarfs,
the initial MSP frequencies are determined by the pre-collapse white
dwarf magnetic field strengths (Yi \& Blackman 1997).

Since we assume that $\Omega_{crit}>\Omega_{dynamo}$, the three classes
are 1) supercritical rotators with strong fields (SPS) 
with $\Omega_{*}>\Omega_{crit}>\Omega_{dynamo}$, 2) subcritical rotators with
strong fields (SBS) with $\Omega_{crit}>\Omega_{*}>\Omega_{dynamo}$, 
and 3) subcritical rotators without dynamo action (SBW) with 
$\Omega_{crit}>\Omega_{dynamo}>\Omega_{*}$. The last class lacks
strong magnetic fields due to the absence of dynamo action.
Since the three classes have very similar rotational frequencies, they
likely originate from very similar pre-collapse initial conditions
of the accreting white dwarf phase (Yi \& Blackman 1997).
Therefore, for an arbitrary distribution of initial white dwarf conditions
(cf. eq. 2-1), the numbers of the three classes of objects are expected to 
be similar.

Because the classes are distinguished by small differences in their spin
frequencies, the electromagnetic dipole luminosity
differs in each class. 
In the two classes, SPS and SBS, the efficient dynamo action results in
$B_*\sim 10^{15}G$, and the dipole luminosity gives
\beq
L_{em}\sim 2\times 10^{50}R_{*,6}^6B_{*,15}^2\Omega_{*,4}^4~ erg/s
\eeq
We expect that SPS's $L_{em}$ ($\propto \Omega_*^4$)
is slightly larger than SBS's due to slightly larger $\Omega_*$ for SPS.
This is apparently consistent with the observed luminosity difference 
between the long bursts and short bursts (e.g. Fishman \& Meegan 1995).
Despite the similar luminosities, the luminosity
evolution time scales are very different. In SPS, the
spin-down, and hence the luminosity decrease, occurs on a time scale
$\sim \tau_{gr}\sim \tau_{gw}\simle 1s$ whereas in SBS, the spin-down and
luminosity decrease occur on a time scale $\sim \tau_{em}\sim 10^2 s$.
We therefore naturally relate SPS to the short bursts and SBS to
long bursts. The third class, SBWs, are expected to have
luminosities (cf. eq. 2-1)
\beq
L_{em}\sim 10^{44} R_{*,6}^6\Omega_{*,4}^{3/2}~ erg/s
\eeq
based on the AIC scenario of Yi \& Blackman (1997). Therefore, unless
there is strong beaming, the third class is not expected to give
rise to cosmological GRBs.

If the critical frequency $\Omega_{crit}$ were 
much lower than the assumed value
(which can happen if the shear viscosity is much lower), the non-axisymmetric
instability could develop at lower $\Omega_*$'s corresponding to
higher $l=m>2$. However, even in this case,
the growth time scale $\tau_{gr}>10^3s$ for $l=m>2$ bar modes
are too long to be of any relevance for AIC produced, hot MSPs. During the
vigorous convection phase, which lasts $\sim 30s$, the $\tau_{em}$ is
shorter than $\tau_{gr}$ for $l=m>2$, which implies that the non-axisymmetry
does not develop before the spin-down occurs through electromagnetic phase. 
Such bursts would resemble those of the SBS class.

If we relate the electromagnetic dipole power to the observed gamma-ray
luminosity assuming $\xi$ as the efficiency to convert the pure electromagnetic
power to gamma-ray luminosity and the physical solid angle of the outflow
is $\Delta\Omega$, we get $\Delta\Omega/4\pi=\xi L_{em}/L_{\gamma}$ where
$L_{\gamma}$ is the observed isotropic gamma-ray luminosity. Therefore,
for $L_{\gamma}\sim 10^{51} erg/s$ and $L_{em}\sim 2\times 10^{50} erg/s$,
we get $\Delta\Omega/4\pi\sim 0.2\times\xi$. For a low efficiency
$\xi\sim 10^{-3}$, the required beam size is as small as $\Delta\Omega
\sim 10^{-3}$. If the outflow's bulk Lorentz factor is $\Gamma$, the physical
beam size need not be smaller than that associated with the relativistic 
beaming unless $\Gamma\simle 30$. 
This bulk Lorentz factor is below the
required value of $\Gamma$ for the gamma-ray transparency
(e.g. Rees 1997), so the jet needs to be only modestly beamed.

\section{Discussion}

The observed bimodal GRB duration distribution has been a mystery.
We have suggested that well-known time scales which are plausible
for the AIC-MSP scenario may account for the bimodality.
Pulsars formed above the gravitationally unstable limit deplete
their energy through gravitational radiation, thereby limiting
the amount of energy available for GRB, and providing the short class.
The long class comes from the gravitationally stable rotators.

Note that this approach to bimodality focuses on 
the GRB engine rather than on the subsequent
nature of how the engine energy is dissipated. 
The latter approach is taken by Sari and Piran (1995) who
suggest that in the context of a (spherical) fireball model,
bimodality may arise from differing roles of Newtonian vs. relativistic 
reverse shocks in energy dissipation. 
Associating the bimodality with the engine rather than the dissipation
assumes that the dissipation mechanism and the efficiency of energy 
extraction into gamma-rays occurs statistically in the same way for the 
two burst populations.  However, we predict a higher luminosity for the
shorter bursts, because the unstable rotators would have the larger 
spin and thus the larger electromagnetic emission before slowing 
down rapidly by gravitational radiation.

An observational signature for associating bimodality directly with
an AIC-MSP source engine would come from the fact that our 
shorter SPS class of GRBs are accompanied by gravitational
radiation which is absent from the longer SBS class.
Provided that young gravitationally radiating 
AIC-MSP deplete their rotational energy through spin-down as described
herein, the detection of gravitational radiation
from short GRBs well in excess of that from long GRBs would give
strong support to the AIC/MSP engine model (Blackman et al. 1996).
The gravitational radiation signature for various evolutionary 
tracks of secularly unstable
AIC pulsars are unique and measurable (Lai \& Shapiro 1995).
In general, gravitational radiation may provide unique signatures
to source engines and be the most promising means of observationally
distinguishing engine models.

If the central engine for GRBs is indeed a rapidly rotating MSP,
the stellar rotation itself may be the source of variability.
The typical MSPs' rotational frequency could naturally give rise
to rapid variabilities on time scales of $\sim 10^{-2}-10^{-3} s$. 
In order for the stellar rotation to show up as variability,
it is required that the magnetic dipole axis is misaligned with
the rotational axis. For large misalignment angles, the amplitudes
of variable fluxes are expected to be large.
If this simple mechanism is indeed responsible for the variability,
duration and variability time scale are not expected to be correlated.
Given the complex burst time profiles (Fishman \& Meegan 1995), 
it is interesting that the
shortest variability time scale is comparable to the shortest GRB duration.
In the MSP scenario, this is naturally explained as the MSP spin-down
time scale can be comparable to MSP spin period.  Note also 
the fact that the GRB engine in this model would remain a stable
pulsar long after the gamma-ray emission. 

If GRBs are well collimated, the long term evolution of the beam is
likely to be influenced by the precession of the MSP jet axis
(Blackman et al. 1996).
Since AIC occurs in a binary system (likely to be a close binary system),
in which the secondary star is a main sequence star with mass
$M_2\sim 0.5M_{\odot}$, the expected precession time scale due to the
Lense-Thirring precession is (e.g. Martin \& Rees 1979)
\beq
P_{LT}\sim {c^2 \over 2(2\pi)^{2/3}G^{2/3}}{(M_*+M_2)^{4/3}\over M_*M_2}
P_{orb}
\eeq
where $P_{orb}$ is the orbital period of the pulsar system and
$M_*\sim 1.4M_{\odot}$. $P_{orb}\sim 1hr$ gives $P_{LT}\sim$ a few $yrs$.
Therefore, if the gamma-ray emission and especially afterglow (on a time
scale $\sim yr$)
occurs within a beam, the precession could give a long term evolution
trend.

The rate of AIC remains uncertain. If the AIC rate is comparable to the
local supernova rate, i.e. $1-10^{-2} yr^{-1}$ per galaxy (Blair 1989), the
observed gamma-ray burst rate of $\sim 10^{-6} yr^{-1}$ per galaxy
is amply explained by the AIC model if $\Delta\Omega\sim 10^{-5}-10^{-3}$
which is largely consistent with the luminosity requirement if
$\xi\sim 10^{-3}$.

The general features of an afterglow from the AIC-MSP scenario may
not be drastically different from that of any other scenario, since an
afterglow represents the dissipation of GRB energy once it has been
produced.  Our focus on the AIC-MSP
model is an attempt to address the physics of the GRB engine source.  
It is important to distinguish the engine physics from that
of the fireball and afterglow in the same way that the physics of an
evolving supernova remnant needs to be distinguished from the physics of
its engine core collapse. Here we have investigated the possibility 
that the bimodal distribution may actually be a signature
for an AIC-MSP GRB engine.

\acknowledgments
I. Y. acknowledges support from SUAM Foundation.

\clearpage

\end{document}